\documentclass [pra,noeprint,reprint,groupedaddress,showpacs]{revtex4-1}
\usepackage{graphicx}
\usepackage{dcolumn}
\usepackage{amsmath}
\usepackage{amsfonts}
\usepackage{amssymb}
\usepackage{bm}
\usepackage{nicefrac}
\usepackage{color}
\usepackage{float}
\usepackage{sidecap}
\graphicspath{{figures/}}
\usepackage{hyperref}
\usepackage[none]{hyphenat}
\hypersetup{backref,
pdfstartview=Fit,
colorlinks=true,
citecolor=blue,
urlcolor = blue,
linkcolor = blue}

\begin{document}
	
\title{Magnon Condensation in a Dense Nitrogen-Vacancy Spin Ensemble}

\author{Haitham A.R. El-Ella}
\affiliation{Department of Physics, Technical University of Denmark, 2800 Kongens Lyngby, Denmark}

\begin{abstract}
The feasibility of creating a Bose-Einstein condensate of magnons using a dense ensemble of nitrogen-vacancy spin defects in diamond is investigated. Through assessing a density-dependent spin exchange interaction strength and the magnetic phase transition temperature ($T_c$) using the Sherrington-Kirkpatrick model, the minimum temperature-dependent concentration for magnetic self-ordering is estimated. For a randomly dispersed spin ensemble, the calculated average exchange constant exceeds the average dipole interaction strengths for concentrations approximately greater than 70 ppm, while $T_c$ is estimated to exceed 10 mK beyond 90 ppm, reaching 300 K at a concentration of approximately 450 ppm. On this basis, the existence of dipole-exchange spin waves and their plane-wave dispersion is postulated and estimated using a semiclassical magnetostatic description. This is discussed along with a $T_c$-based estimate of the four-magnon scattering rate, which indicates magnons and their condensation may be detectable in thin films for concentrations greater than 90 ppm.  
\end{abstract}
\maketitle

\section{Introduction}

	\begin{figure*}
	\centering
	\includegraphics[scale = 0.95]{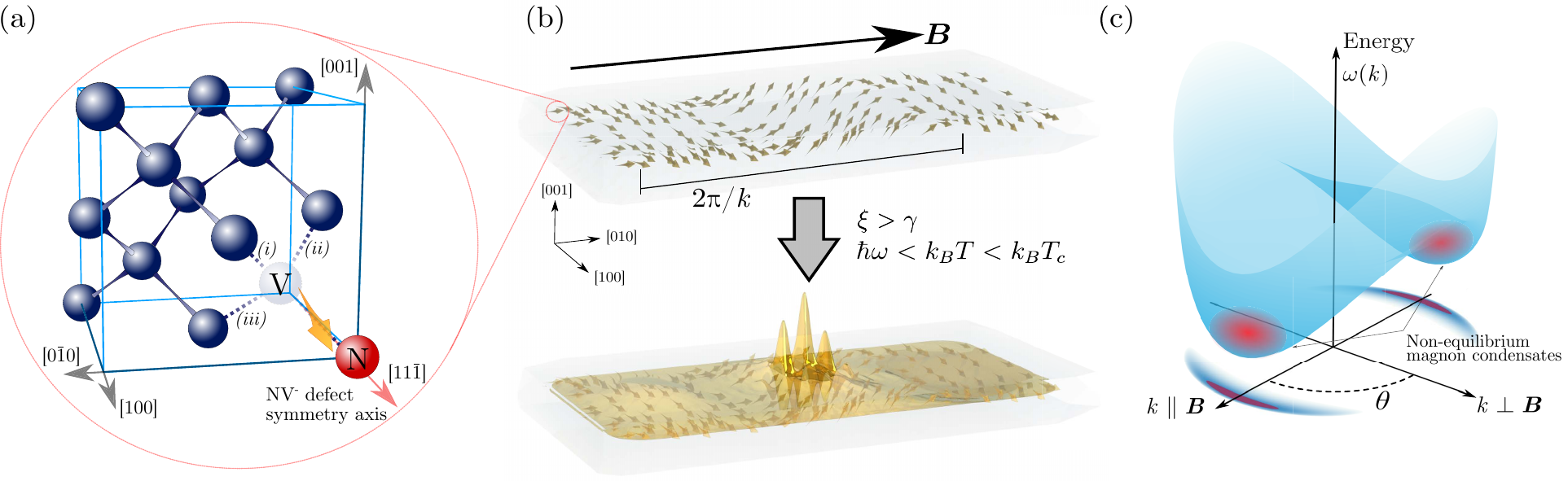}
	\caption{(a) Schematic of a diamond unit cell, highlighting the configuration of the N-$V^-$ defect and its three other possible orientations [(\textit{i})-(\textit{iii})] relative to the crystal coordinates. (b) Magnons will condense into their minimum energy $\hbar\omega$ through inelastic scattering at a rate $\xi$ provided that this rate is greater then the spin-lattice relaxation rate $\gamma$, and that $\hbar\omega$ is less than the local thermal energy $k_B T$ and the magnetic phase transition energy $k_BT_c$. (c) A surface plot of the dispersion of dipole-exchange plane spin-waves for all relative angles between an external magnetic field $\mathbf{B}$ and the propagating spin wavenumber $k$. Two global minima occur in the dispersion of spin waves propagating in parallel with $\mathbf{B}$.}
	\end{figure*}  		

	Bose-Einstein condensates are fascinating systems which tangibly demonstrate macroscopic quantum coherence and novel matter-wave dynamics. Following their initial demonstration using ultracold atom gases \cite{Davis1995, Anderson1995}, the phenomenon has been observed in solid systems using metastable ensembles of photonic \cite{Kasprzak2006a,Plumhof2014} and magnetic \cite{Nikuni2000, Demokritov2006} quasi-particles. Here, the possibility of observing similar magnetic quasiparticle condensation in an ensemble of charged nitrogen vacancy (N-$V^-$) spins in diamond is investigated.  
		
	The pursuit of solid-state based condensates has been motivated mainly by the desire to observe the phenomenon at less stringent temperatures and critical densities \cite{Imamoglu1996}. This is facilitated with the use of quasi-particles due to their orders of magnitude lower mass compared to atoms. However, this advantage is mitigated by their short lifetimes, which is limited by the local density of states and lattice-induced relaxation. Such condensates are understood to be formed by being in-elastically scattered into their lowest energy at a critical density, provided that the scattering rate is faster than their lifetime, which can be limited by both intrinsic and extrinsic factors. Quasi-particle condensates are therefore transient by nature and are often referred to as non-equilibrium condensates, due to not being thermally equilibrated with their immediate environment. In photonic systems, most reports have focused on the condensation of cavity-based exciton-polaritons \cite{Byrnes2014} which possess intrinsic lifetimes in the order of hundreds of picoseconds \cite{Deng2010}, being usually limited by the quality of the optical cavity. In magnetic systems, most observation of condensates have been through exploring deviations in the magnetic and temperature dependent phase transitions of quantum magnets \cite{Zapf2014}. However, recent observations have centred around the condensation of magnons in thin-film yttrium iron garnet (YIG), which are limited by their spin-lattice relaxation time in the order of hundreds of nanoseconds \cite{Serga2014}. 

	Magnons, or spin-wave quanta, constitute propagating disturbances in a correlated array of magnetic spins \cite{Bloch1930,Stancil2009}, and whose modes and dynamics are being intensely studied in a variety of systems in the pursuit of both fundamental many-body states \cite{Sun2017}, and practical information technology \cite{Karenowska2015,Chumak2017}. While the manifestation of spin-waves in magnetic media is ubiquitous, their room-temperature quantum condensation is a recently observed phenomenon, and presents a compelling platform for exploring many-body and macroscopic quantum dynamics at room temperature \cite{Plumhof2014, Sun2017a, Ballarini2017, Lerario2017}. Given the central role of spin-lattice relaxation, it is compelling to consider their manifestation and condensation in magnetically doped crystals with long-lived spin systems, and for which additional degrees of spin manipulation may be available.   
	
	The interest in exploring N-$V^-$ ensembles is therefore motivated by their record room-temperature spin-lattice relaxation times \cite{Balasubramanian2009,Jarmola2012}, and the demonstrated high fidelity of coherent control and optical readout from both single spins and ensembles, without the need for cryogenics or intricate detection schemes \cite{Doherty2013}. In light of their exceptional spin properties, N-$V^-$ ensembles have been suggested for the study of controllable quantum spin-glasses \cite{Lemeshko2013} and cavity-mediated ferromagnetic ordering \cite{Wei2015}, and have already proven to be experimentally versatile systems for developing quantum-based metrology and information protocols \cite{Wu2010,Barry2016a,Greiner2016}, room temperature masers \cite{Breeze2018}, and the exploration of many-body quantum systems \cite{Yao2010}, demonstrated in particular by the recent observation of time crystals \cite{Choi2017}.
		
	Fundamentally, condensation of quasi-particles is dependent on the presence of a global minima or `traps' in their momentum space, and an inelastic scattering mechanism ($\xi$) which enables a form of evaporative-cooling for condensation to occur. This needs to occur at a rate which exceeds the spin-lattice relaxation ($\gamma$), and towards a minimum energy that is below the surrounding thermal energy. In turn, the thermally equilibrated energy of the crystal matrix needs to be lower than the magnetic phase transition temperature ($T_c$, Fig.1), being the point where the interaction energy between nearest-neighbour spins exceeds the local thermal energy. This ensures that an ensemble of magnetic spins will mutually lock each other in phase, bringing about long-range magnetic order, and accommodating the propagation of magnetic order perturbations, i.e. spin-waves \cite{Stancil2009}. 
	
	To evaluate these points, the density-dependent spin-exchange and dipole interaction strengths in a randomly distributed ensemble of N-$V^-$ spins is assessed. This is carried out with the understanding that the spin properties may become unpredictably modified when defect concentrations begin approaching the crystal unit cell density, thereby limiting this analysis to densities that retain the spin defects integrity, i.e. that defects are separated enough throughout the crystal lattice to ensure that their electronic level structure is not fundamentally modified. There are considerable difficulties in generating dense N-$V^-$ ensembles with concentrations exceeding 10 ppm, and for concentrations reported in the literature, almost all do not exceed a few ppm. Exceptionally, there are reports of concentrations exceeding 10 ppm and approaching 100 ppm \cite{Mrozek2015,Kucsko2018,Giri2018}. This encourages the consideration of higher concentrations, and the optimistic outlook that there are no intrinsic limitations, as growth optimisation techniques and N-$V^-$ creation efficiencies are continuously improved. 
		
	This article is divided into two sections. In the first section, a general consideration of the N-$V^-$ density limit is outlined, followed by an estimation of the exchange interaction strength $J_{ex}$. This is based on first principles, and is delineated as a function of concentration, which is then parameterised in terms of the average nearest-neighbour distance. The obtained values are then applied to estimate the magnetic phase transition temperature $T_c$ as a function of concentration. In particular, the Sherrington-Kirkpatrick model is used \cite{Sherrington1975} in combination with the average exchange interaction $\langle J_{ex}\rangle$, under the application of a magnetic field. In the second section, the calculated concentration-based $\langle J_{ex}\rangle$ and $T_c$ values are used to justify the use of a semi-classical description of the dipole-exchange plane spin-wave dispersion \cite{Kalinikos1990}, as well as the rudimentary condensation dynamics based on the analysis formulated in \cite{Serga2014}. As previously mentioned, observation of room-temperature magnon condensation has been extensively reported so far for thin-film YIG, which thereby provides an instructive physical comparison. Its reported parameters are therefore highlighted in section III, in order to contextualise the presented values. 
	
\section{Magnetic Order in an N-$V^-$ Spin Ensembles}
			
	\subsection{Density Considerations}	
	
		\begin{figure}
		\includegraphics[scale = 0.5]{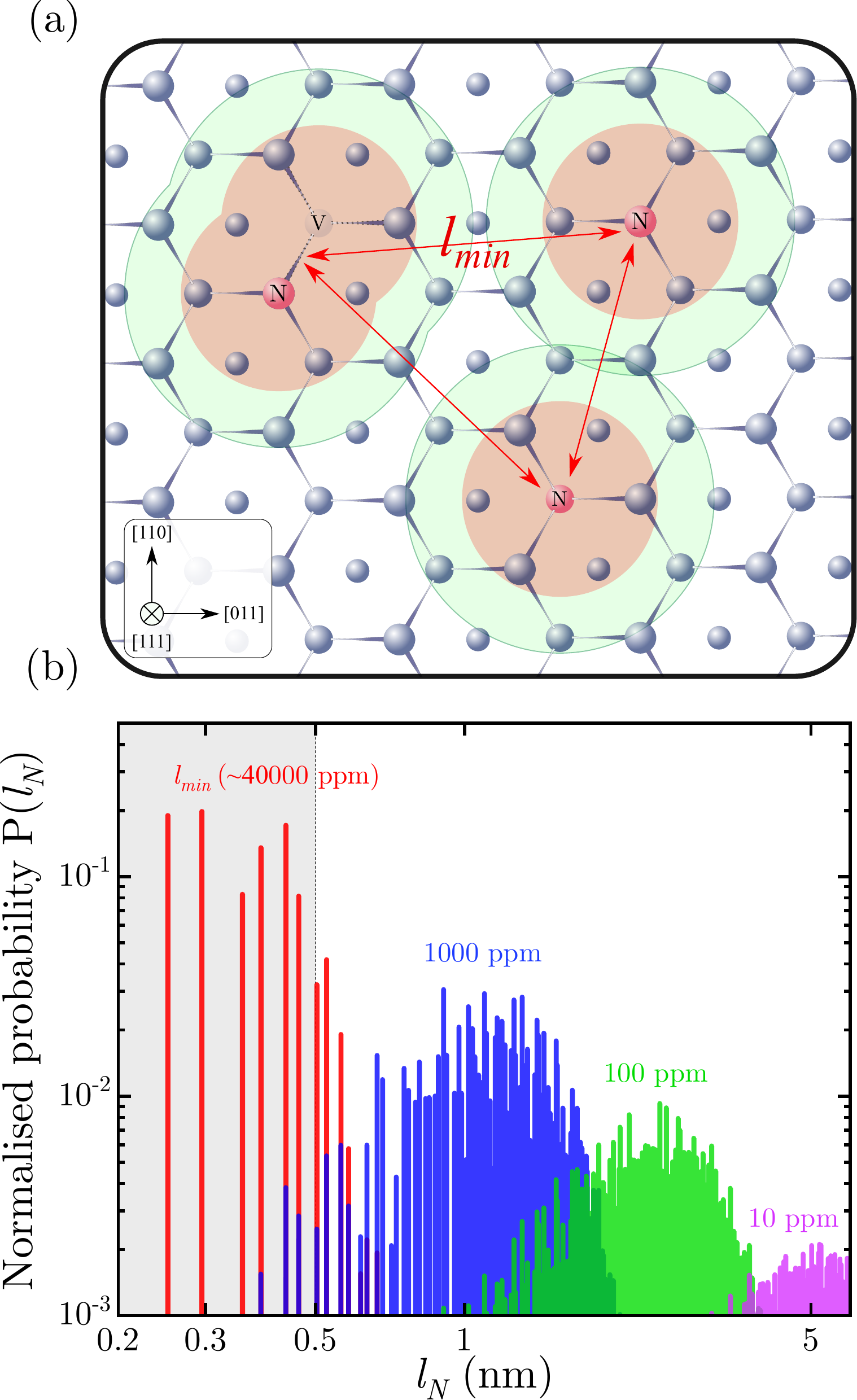}
		\centering
		\caption{(a) Simplified crystal schematic highlighting the considered 26 lattice-site cluster surrounding a N-$V^-$ defect, and their minimal distance in the order of approximately 5 \AA~if they do not fully overlap (red and green circles show first and second coordination shells, respectively). (b) Plots of the probability-density equation (\ref{1}), for four different concentrations, highlighting the concept that for a random distribution at concentrations exceeding 1000 ppm, a considerable portion of the ensemble will occupy nearest-neighbour lattice sites that are shorter than the conceptual 5 \AA ~limit discussed in the text.}  
	\end{figure}	

		\begin{figure*}
		\includegraphics[scale = 0.9]{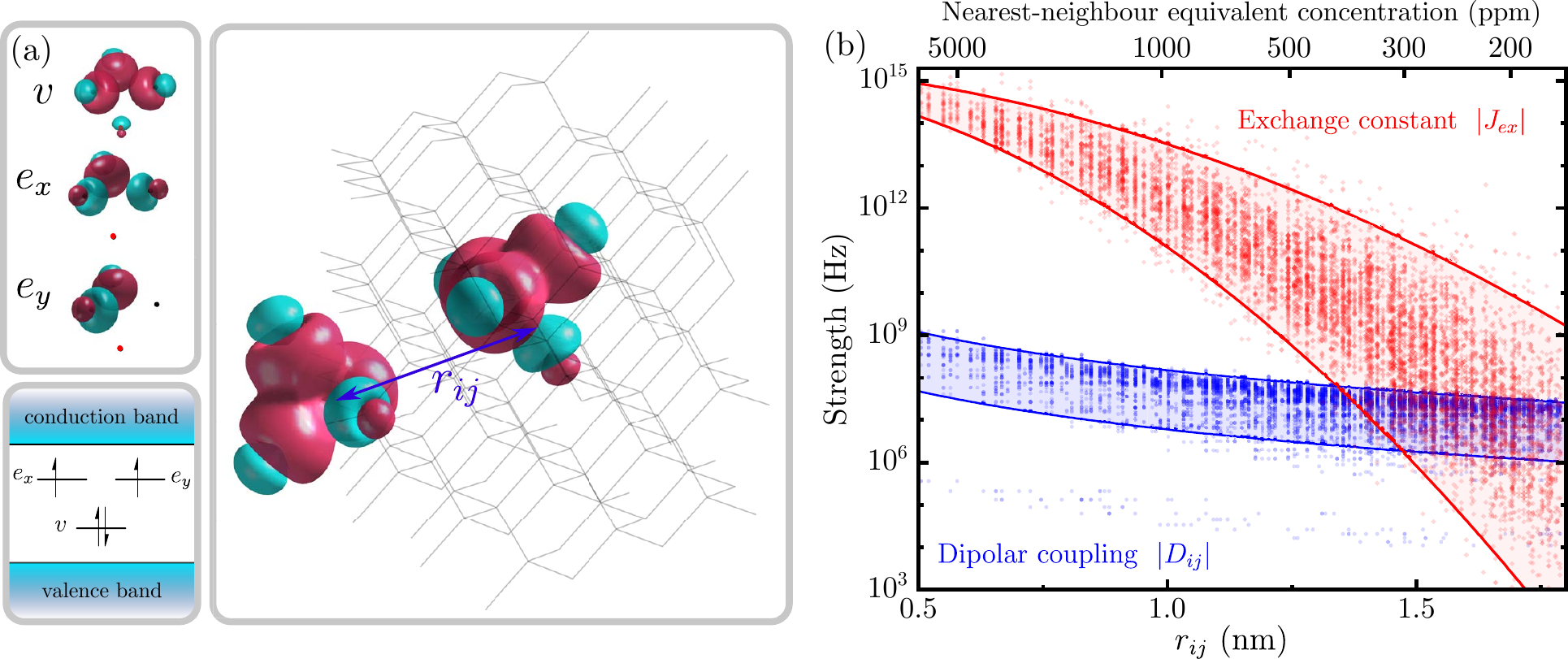}
		\centering
		\caption{(a) Isosurface plots (5\% of peak amplitude) for the three wavefunctions of the N-$V^-$ states that lie within the diamond bandgap, along with one possible arrangement in a diamond lattice (shown using the $v$ orbital). Red/turquoise highlight regions of opposite phase, while the black/red dots highlight the carbon/nitrogen atom positions. (b) The dipole and exchange energies plotted as a function of the possible separation and relative orientation within the diamond crystal lattice. Each $J_{ex}$ value is bound by a $\pm18$\% error bar, which is not shown for clarity. The shaded regions highlight where at least 95\% of the calculated values are spread for given separation distances.}
	\end{figure*}	

		Defects in a crystal induce lattice distortions which are amplified by the density of neighbouring defects. This may lead to the energetically favourable formation of aggregates \cite{Deak2014}, and a limit on the obtainable defect density based solely on the relaxation of lattice-strain is conceivable. In diamond, the distortion introduced by substitutional nitrogen or vacancy defects has been calculated to displace the nearest neighbour carbon atoms by approximately 1\% \cite{Prentice2017}, and reported experimental observations have indicated that lattice distortions surrounding a N-$V^-$ centre are negligible beyond approximately $4$ \AA ~\cite{Abtew2011, Alkauskas2014}. In addition to strain, the integrity of a defects electronic level structure can be assessed through the spatial density of its spin wavefunction. For substitutional defects in diamond such as Ni or Cr, the related spin wavefunctions do not extend much beyond the first neighbouring carbon atoms, and they may therefore be considered as a spherical cluster of five lattice sites, which retains the defects integrity provided that the nearest neighbour is three lattice-sites away. In the case of twin point defects such as the N-$V^-$ centre, the resulting wavefunction of the spin states are non-centrosymmetric, and the volume needed to retain their integrity is much less obvious. The N-$V^-$ centre - a pair of point defects (a substitutional nitrogen and a carbon vacancy), for which the symmetry axis ($C_{3v}$) pins the quantisation axis, as shown in Fig.1(a) - is understood from \textit{ab initio} supercell and tight-binding models \cite{Gali2008, Kortan2016} to possess a spin wavefunction that is centralised around the vacancy and the dangling bonds of the three surrounding carbon atoms. Accounting for these properties, a self-contained N-$V^-$ defect is conservatively considered here as a cluster of 26 lattice sites. This consideration implies a maximum limit to the density of N-$V^-$ of one defect pair per 26 lattice sites, corresponding to a concentration of approximately $4 \times 10^4$ ppm. If these clusters are considered to be ``hard-spheres", such that they do not overlap, as conceptualised in Fig.2(a), then their tightly packed arrangement would imply a nearest-neighbour distance of approximately 5 \AA, (or about one defect for every three unit cells). However, unless specific strategies are implemented to facilitate a degree of order during defect creation (e.g.\cite{Bayn2015, Ishiwata2017}), an ensembles nearest-neighbour separation distances will be normally distributed. This implies that a lower practical limit exists for ensuring that the probability of defect aggregation is negligible.
		
		A rudimentary estimate of a practical concentration limit can be carried out by considering the nearest-neighbour distance from a single point in a crystal lattice. This can be described in terms of the defect density $N_d$ and the number of possible sites at a distance $l_N$ \cite{Chandrasekhar1943, Hall2014}, leading to the probability-density expression $P(l_N)$, and its first moment:
		\begin{align}
		\label{1}
		P(l_N)&=(4\pi l_N^2 N_d)e^{-(4\pi l_N^3 N_d)/3},\\
		\label{2}
		\langle l_N \rangle &= \bigg(\frac{3}{4\pi N_d}\bigg)^{1/3}\Gamma\bigg(\frac{4}{3}\bigg),
		\end{align}  
		where $\Gamma$ is the gamma function. Accounting only for the allowed relative positions within the diamond lattice, $l_N$ is discretized and Eq.(\ref{1}) is appropriately normalised to obtain a binomial probability distribution as a function of $N_d$, as shown in Fig.2(b) for a range of concentrations. Using this representation, and setting a 5\% cumulative probability threshold on the formation of aggregates (i.e. for the probability of the nearest-neighbour distance being less than approximately 5 $\mathring{\text{A}}$), a defect integrity-retaining concentration limit is estimated to be approximately $N_d/N_c \approx 1100$ ppm (where $N_c$ is the carbon atom density). Although somewhat arbitrary, this limit is chosen to provide a grounded basis for the subsequent estimations. This is considered bearing in mind that the long-range structural and electronic perturbations that could occur for concentrations approaching this value may further decrease this limit, especially when considering the presence of neighbouring non-N-$V^-$ defects, such as $^{14/15}$N and $^{13}$C atoms. For example, in light of the N-$V^-$ systems unique charge-state dynamics \cite{Giri2018}, a finite concentration of electron donors, such as the  $^{14/15}$N defect, is necessary to ensure that the N-$V^-$ $\rightarrow$ N-$V^0$ conversion rate is negligible. Therefore, their relative density may also encourage the formation of aggregates and further limit the electronic integrity-retaining concentration. At the least, their concentration should not exceed that of the N-$V^-$ ensemble's, considering similar nearest-neighbour arguments.
		
	\subsection{Spin-Spin Interactions}

		Isolated magnetic impurities interact through both long-range dipole coupling and short-range spin exchange coupling. In the presence of a magnetic field, the dipole interaction can be described using the secular approximation, in terms of the separation and relative angle of two spin dipoles pointing along the magnetic field. The spin exchange interaction consists of Coulomb ($C_{ij}$) and spin-exchange ($X_{ij}$) energies \cite{Stancil2009} which are a function of the N-$V^-$ spin wavefunction $\psi_{i}(r_i)$. The resulting exchange coupling energy ($E_{ex}$) is the normalised sum or difference of these two energies, for a singlet or triplet configuration, respectively:     
		\begin{align}
		\label{3}
		D_{ij} &=\frac{h\mu_0\gamma_e^2}{4\pi}\frac{1-3\cos{\theta_{ij}}}{r_{ij}^{3}},\\
		\label{4}
		E_{ex}^{\pm} &= \frac{C_{ij} \pm X_{ij}}{1 \pm \alpha_{ij}^2},\\
		\label{5}
		C_{ij} &= \iint \vert\psi_{i}(r_i)\vert^2\frac{e^2}{4\pi\varepsilon_0 r_{ij}}\vert\psi_{j}(r_j)\vert^2dV_idV_j,\\
		\label{6}
		X_{ij} &= \iint \psi_{i}^*(r_i)\psi_{j}^*(r_j)\frac{e^2}{4\pi\varepsilon_0 r_{ij}}\psi_{i}(r_j)\psi_{j}(r_i)dV_idV_j,
		\end{align}
		\begin{align}
		\label{7}
		\alpha_{ij} &= \int \psi_{i}^*(r_i)\psi_{j}(r_j)dV,
		\end{align}	
		where $r_{ij} = r_{i} - r_{j}$ is the distance and $\theta_{ij}$ is the angle between neighbouring spins pointing along an externally applied magnetic field, $\gamma_e$ is the electron gyromagnetic ratio, $e$ is the elementary charge constant, and $\mu_0,~\varepsilon_0$, are the vacuum permeability and permittivity, respectively. The wave-functions $\psi(r)$ are described here using a linear combination of $sp^3$ hybridised orbitals which are constructed using the hydrogenlike orbitals for the C and N atoms outer electron shell \cite{Lenef1996, Hossain2008} (see Appendix \ref{A} for the full expressions used). The iso-surfaces of the three N-$V^-$ orbitals that exist within the diamond band-gap are shown in Fig.3(a). While $E_{ex}$ is the resulting energy for spin orbital overlap and electron repulsion, the metric describing the collective ordering of spins is the energy needed for shifting between a singlet and triplet configuration $J_{ex}= E_{ex}^{+} - E_{ex}^{-}$ \cite{Stancil2009}. 
		
		The expressions $\vert D_{ij}\vert $ and $\vert J_{ex}\vert $ are plotted in Fig.3(b) as a function of the allowed relative positions from a central N$V^-$ spin. A binomial distribution of values is shown for each allowed $r_{ij}$, which widens as the density decreases and the nearest-neighbour distance spans more lattice positions. The shaded region, bounded by their characteristic fitting functions, highlights 95\% of the calculated values. For clarity, only values calculated considering two $e_x$ orbitals are plotted, and a similar spread of energies is obtained for $e_x$-$e_y$ and $e_y$-$e_y$ combinations. The volume integrals in Eq.(\ref{5})-(\ref{7}) are numerically solved using Monte-Carlo integration and are error-bound by approximately $\pm 18 \%$. An $r^{-3}$ dependence is fitted for $\vert D_{ij}\vert $, while for $\vert J_{ex}\vert $ a $e^{-r^2}/r$ is used. 
		
		The calculated values here show that the average absolute exchange constant $\langle\vert J_{ex}\vert\rangle$ exceeds $\langle\vert D_{ij}\vert\rangle$ by up to six orders of magnitude as $r_{ij}$ approaches 0.5 nm. However, unlike $J_{ex}$, the spread of $D_{ij}$ for a random distribution is centred around zero, due to the angular cosine dependence in Eq.(\ref{3}). The average dipole strength $\langle D_{ij}\rangle$ therefore approaches zero as the density is decreased, while $\langle J_{ex}\rangle > \langle D_{ij}\rangle$ when $r_{ij}<2.4$ nm, corresponding to concentrations exceeding approximately 70 ppm. It also highlights, as expected, the quicker decay of $\langle J_{ex}\rangle$, and how the relative spread of energies drastically increases by several orders of magnitude. 
		
		While the $sp^3$-hybrid orbital basis used here for estimating $J_{ex}$ is rudimentary compared to more sophisticated electron structure calculations, this approach has been previously employed \cite{Ho2016}, and the estimated values show an order of magnitude and trend that is consistent with values obtained using a tight-binding basis \cite{Kortan2016, Chou2018}.
		
	\subsection{Magnetic Phase Transitions}
	
		The average and variance of the calculated interaction strengths provides a basis to parameterise them as a function of the average nearest-neighbour distance $\langle l_N\rangle$, and thereby the corresponding density $N_d$, through Eq.(\ref{2}). These values allow for the determination of the collective magnetic order, while accounting for the systems inherent randomness. The study of magnetic disorder is central to the study of real systems, for which a fully consistent three-dimensional theoretical picture has yet to be developed, due to the intractability of considering their thermodynamic limit \cite{Binder1986}. One much-studied formalism which has been reasonably successful in treating random systems is the replica-symmetric approach for solving the Sherrington-Kirkpatrick (SK) Hamiltonian \cite{Sherrington1975} in the presence of a random field:
		\begin{align}
		\label{8}
		\hat{\mathcal{H}} &= -\frac{1}{2}\sum^{N_dV}_{\langle i,j\rangle}W_{ij}S_iS_j -\gamma_e \sum^{N_dV}_{i} h_i S_i,
		\end{align}	
		where $S_i = \pm1$ are Ising spin variables, $\{W_{ij}\}$ is the set of random interactions between nearest-neighbour pairs, and $\{h_i\}$ is the set of random magnetic field strengths. The angled bracket indicates summation only over nearest neighbour pairs, and the number of spins is defined in terms of the considered density $N_d$ and volume $V$. Using the replica-symmetric approach \cite{Parisi1983,Soares1994,Hadjiagapiou2014,Magalhaes2011}, a heuristic solution for Eq.(\ref{8}) is derived for the systems free energy and magnetic order parameters, and thereby the magnetic phase transition temperatures from paramagnetic to ferromagnetic order $T_{c}$. Without delineating the derivation steps and transformations, (see Appendix \ref{B}), the solved expressions assume a normal distribution of $\{W_{ij}\}$ and $\{h_{i}\}$, and consists of coupled expressions in terms of the normalised variance and average of the total interaction strengths $\{W^2,W_0\}$ and the magnetic field $\{\Delta^2,h_0\}$ : 
		\begin{align}
		\label{9}
		T_{c} &= W_0\int^{\infty}_{-\infty}\frac{e^{-z^2/2}dz}{\sqrt{2\pi}}\text{sech}^2[\zeta-\beta(h_0-W_0m)],\\\nonumber
		\zeta &= \beta \big[h_0 + W_0 m + z W \sqrt{q + (\Delta/W)^2}\big], \nonumber 
		\end{align}			  		
		where $m$ is the magnetic order parameter, $q$ describes the long-time magnetic correlations (sometimes referred to as the spin-glass order parameter), the integral $\int f(z) dz$ arises from the Hubbard-Stratonovich transformation (an exponential identity applied during derivation), and $\beta = 1/k_B T$, $k_B$ being Boltzmans constant and $T$ being the local temperature.
		
		Given that the SK Hamiltonian and the derivation for $T_c$ is defined in terms of Ising spin variables, while the N-$V^-$ is distinctly a spin-1 system, a simplification is necessary to justify the use of the SK model. Furthermore, as highlighted in Fig.1(a), the quantisation axis of the N-$V^-$ centre may be pinned along four different $\langle111\rangle$ directions, implying that for a randomly distributed ensemble, there can be four subgroups projected at differing angles along an externally applied magnetic field. Therefore, a static external magnetic field is applied parallel to either the [010] crystallographic axis, such that each possible N-$V^-$ orientation is equally projected onto this external field, or along one of the [111] quantisation axis which results in one fully parallel projected group, and three equally acutely projected subgroups to the external field. 
		
		The presence of a non-axial magnetic field modifies the eigenstates into superposition of the spin-1 states, $\vert\varphi^{\pm,0}\rangle$, and the spin quantum number $m_s$ ceases to be a good one. This implies violation of the magnetic selection rules, allowing for normally forbidden $\Delta m_s = \pm2$ transitions. However, these transitions will be relatively suppressed compared to the allowed transitions beyond moderate applied magnetic field strengths, while the allowed transitions (highlighted by the black arrows in Fig.4(a))  will gradually be energetically equivalent. Therefore, to further simplify the spin-1 system, either a field strength of 102.4 mT is applied to bring about degeneracy of the $\vert 0 \rangle$ and $\lvert-1\rangle$ eigenstates for one of the subgroups, or a field strength beyond 180 mT is applied to energetically equalise the two allowed $\vert\varphi^{\pm}\rangle\leftrightarrow\vert\varphi^{0}\rangle$ transitions such that only two spin-state superpositions are considered to exist. In the former case, the remaining subgroups are ignored as their spin-state become ill-defined, and in the latter case, the $\Delta m = \pm2$ transition is ignored as its likelihood is relatively  suppressed as a function of the applied field strength.  

		These simplifications for degeneracy are not strictly applicable - the spin states of the N-$V^-$ system are further split into hyperfine lines (two or three for $^{15}$N or $^{14}$N, respectively) due to the interaction of the defects electron and nitrogen nuclear spins, and the application of a non-axial magnetic field will bring about spin-mixing and an avoided crossing of the transitions at their degeneracy point \cite{Broadway2016}. Nonetheless, this fact may be side-stepped by considering the inherent in-homogeneous broadening of dense spin ensemble resonance energies \cite{Acosta2009}, and the spread of interaction energies shown in Fig.3(b), which will exceed the resulting anti-crossing splitting energy. Under these conditions, the states can be well-approximated as being degenerate for the analysis applied here.
		
		The concentration-dependent mean and variance are extracted from the analysis in the previous section. A normal approximation of the binomial distribution of values for each allowed separation distance is used, and related to a given concentration through Eq.(\ref{2}). The extracted values are combined such that $W_0 = \langle J_{ex}\rangle + \langle D_{ij}\rangle$, and setting $h_0 = 180$ mT, a rudimentary phase diagram is calculated and plotted in Fig.4(b) as a function of concentration. The plot highlights a first-order phase transition between a paramagnetic and ferromagnetic phase, for which the order parameters shift from $\{m,q\}=0$ to $\{m,q\}\neq 0$. A smaller region at low temperatures highlights a spin-glass phase for which $\{m=0,~q\neq0\}$, but which reverts to a ferromagnetic phase in the presence of an external field ($m\neq0$ when $h_0>0$). The shaded region highlights the uncertainty of the $T_c$ point, based on the uncertainty of the estimated $\langle J_{ex}\rangle$ values. The first-order phase transition implies a spontaneous ferromagnetic ordering of the spin ensemble into either its collective $|1\rangle$ state or its approximately degenerate $\vert-1\rangle$ and $|0\rangle$ states. The phase diagram shows that $T_c$ reaches room temperatures when the concentration is approximately $450\pm100$ ppm, while it becomes experimentally accessible ($T_c>10$ mK) when the concentration exceeds approximately 80 ppm. 
		
					\begin{figure}
			\centering
			\includegraphics[scale = 0.42]{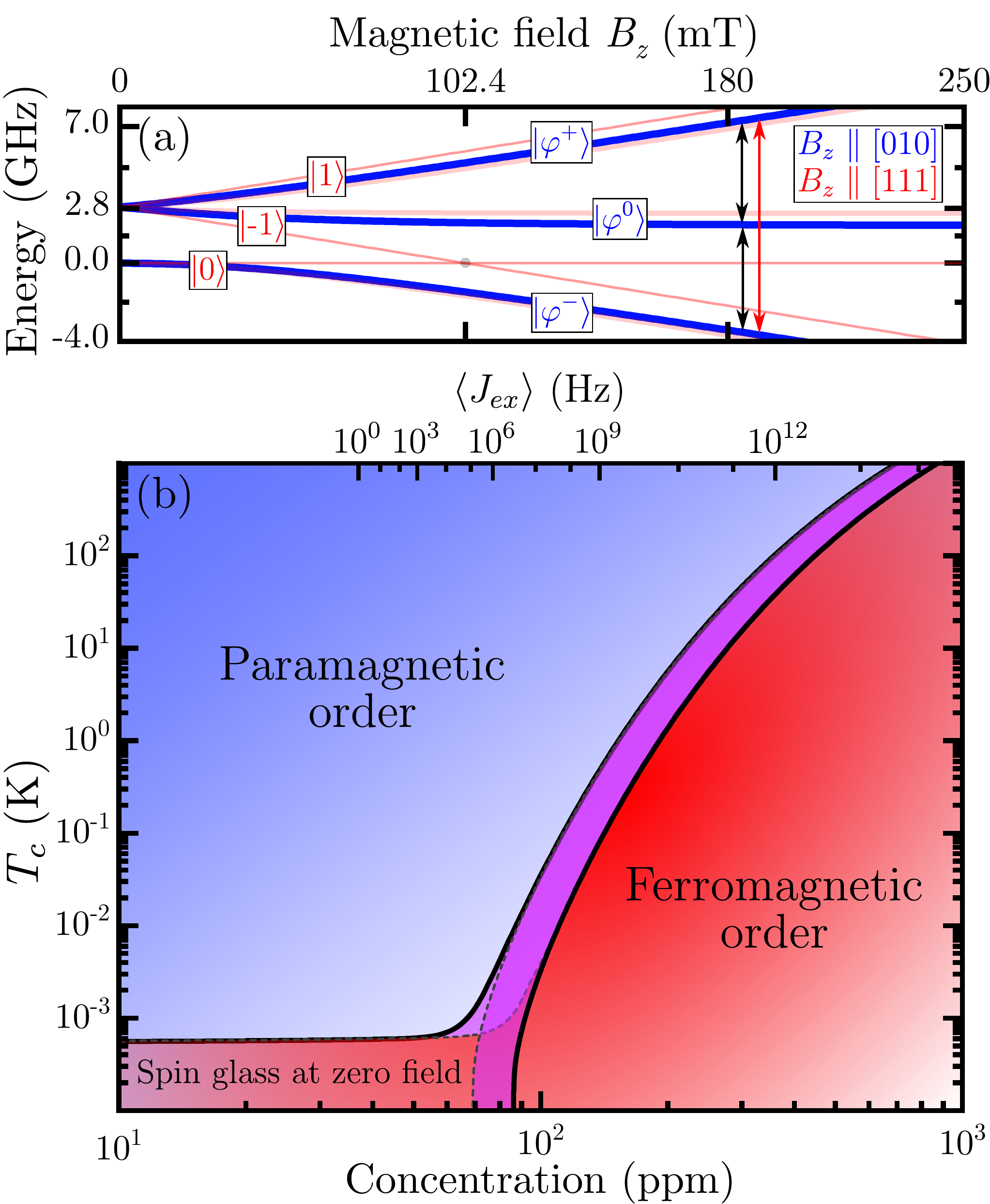}
			\caption{(a) Zeeman splitting of a N$V^-$ ensembles three spin eigenstates shown for two different magnetic field orientations. When aligned along [010], as opposed to [111], all four subgroups symmetry axes are offset by the same angle, and their eigenstates are therefore degenerate. At field strengths beyond approximately 180 mT, the allowed transitions between the resulting mixed-spin levels (black arrows) become near-equal, while the disallowed transition (red arrow) becomes relatively suppressed. At these fields, or at the ground-state anticrossing inducing field (grey dot), the total system may be described in terms of two possible spin state. (b) Phase diagram of the magnetic order as a function of concentration, using the average and variance of $D_{ij}$ and $J_{ex}$ as estimated in Sec. II.B. The purple shaded region highlights the uncertainty based on the error-bound of $\langle J_{ex} \rangle$, while the values shown in the top axis are similarly error-bound.}
		\end{figure}  		
		
		The noise in the interaction energies, and their effect on $T_c$, is encapsulated in the SK model though accounting for the interaction energy variance $W^2$. Further electron spin noise stemming from interactions with the electron spins of $^{14/15}$N substitutional impurities should be of the same order, provided their concentration does not exceed that of the N-$V^-$ spin ensemble. The presence of a $^{14/15}$N and $^{13}$C nuclear spin bath may also be considered in terms of their hyperfine coupling and the magnetic spin noise they introduce. However, these are comparably negligible as the nuclear-electron spin hyperfine coupling is, at most, approximately 130 MHz for $^{13}$C and N-$V^-$ spins in adjacent lattice sites \cite{Felton2009}, and the effective magnetic field strength they introduce will be orders of magnitude smaller than the variance $W^2$, for concentrations exceeding 80 ppm. Another consideration is the effect spin noise will have on breaking the enforced degeneracy of the $\lvert-1\rangle$ and $|0\rangle$ states. Provided that the surrounding magnetic spin noise does not exceed the inherent noise set by the interaction strength variance, this should not affect the first-order phase transition temperature derived here. This should hold true if the surrounding concentration of electron spins does not exceed that of the N-$V^-$ ensemble. 
		
		In delineating the magnetic phase map, the onset of long-range magnetic order is highlighted, allowing for the assumption of spin-wave manifestation and propagation below $T_c$ for any given concentration, in the presence of an external magnetic field. Fundamentally, the collective strength of the exchange interaction must exceed the local thermal energy, such that $W_0 (N_d V) > k_B T$, in order to preserve phase-correlated magnetic disturbances, and ensure spin-wave propagation. With these obtained parameters, the description of spin-waves may be carried out beyond this point using the conventional picture for dipole-exchange plane spin-waves, without the further need to invoke the particular characteristics of the N-$V^-$ system. 
		
		It is worth stressing that the SK Hamiltonian in Eq.(\ref{8}) is not used as a basis to calculate the possible spin-wave dispersion in the subsequent sections. Ising variables do not represent continuous spin operators which are fundamental for describing spin-wave characteristic. The SK model is employed here because of the convenience of the resulting expression, Eq.(\ref{9}), for $T_c$, which accounts for a distribution of interaction and magnetic field values in the thermodynamic limit while being relatively easily solved numerically. More sophisticated Heisenberg-based models are warranted, however, considering the application of a non-zero symmetry-breaking external field and the modification of spin states to resemble that of a spin-half system, the calculated $T_c$ using a mean-field-based Heisenberg model is expected to give similar results for the first-order phase transition. In fact, Ising-like models can be defined as approximately equivalent to Heisenberg-like models under a mean-field approximation for time-averaged values of nearest-neighbour spins. In light of this, their approximate correspondence under the conditions described here should be sound, in particular for predicting a first-order ferromagnetic transition. These considerations reflect reported comparisons of these two models, for example when predicting tricritical points considering a random field distribution \cite{Santos-Filho2016} or comparing both models to measured $T_c$ values of disordered alloys \cite{AguirreContreras2005}.   
		
		The heuristic solution of the SK model for estimating $T_c$ is therefore taken to be a good approximation for predicting when a first-order magnetic phase transition may occur under the specific conditions stated here, and as a justification for the assumption of spin-waves manifestation for $T_c$-related concentrations.

\section{Spin Wave Condensation}
	
	\subsection{Magnon Dispersion}
				
		A similar picture for the dispersion of plane spin-waves can be obtained using either a quantum \cite{Kreisel2009} or semi-classical framework, for which a thorough exposition can be found in \cite{Capolino2009}. Here, a classical approach is used, as described in \cite{Kalinikos1986a, Kalinikos1980, Capolino2009}, for a layer with a finite thickness and infinite length and width. The exact dispersion relations can be obtained from a system of Maxwell and linearized Landau-Lifshitz equations of motion, for which the solution consists of an infinite set of vector amplitude functions whose vanishing determinant gives the exact dispersion relation for plane spin-waves. This is defined as a function of thickness $d$, the exchange field stiffness constant $\eta$, the saturation magnetisation $\omega_M$, and the relative angles of the wave propagation vector and the externally applied magnetic field $\theta_B$, and $\phi_B$, respectively. 
				
		For an ensemble of magnetic spins, the saturating magnetisation $\omega_M$ can be approximated as a function of the number of spins in a given volume such that $\omega_M \approx \gamma_e 4\pi g \mu_B N_dV$, where $g$ is the electron $g$-factor and $\mu_B$ is the Bohr magneton. The exchange field stiffness constant $\eta$ is a parameter conventionally used to describe the exchange coupling strength in ferromagnetic material with respect to their crystal structure \cite{Stancil2009}. Given that the ensemble is randomly distributed and therefore possesses no long-range symmetry, this is defined here in terms of Eq.(\ref{2}): 
		\begin{align}
		\label{10}
		\eta &\approx \frac{4\pi}{9 \Gamma(\tfrac{4}{3})^3}\frac{\langle J_{ex}\rangle \langle l_{N}\rangle^5}{\mu_0 \mu_B^2g^2}.
		\end{align}
		The exchange field stiffness values obtained mirror the concentration dependent trend shown in Figs.3(b) and 4(b), which may be considered in light of the fixed constants of the YIG system \cite{Stancil2009}. For YIG, the reported exchange field stiffness constant is approximately $3\times10^{-16}$ m$^{-2}$ (or approximately 30 GHz in terms of $T_c k_B/h $). The N-$V^-$ ensemble exceeds these values for concentrations beyond 80 ppm, while the ensembles saturation magnetisation will always be at least three orders of magnitude lower than that reported for YIG at approximately 175 mT, for the N-$V^-$ concentrations considered here.		
		
		With the assumption of ferromagnetic order, the description of plane spin-waves and their dispersion is carried out using the approximate form first described in \cite{Kalinikos1986a}, for isotropic ferromagnetic layers. Although it has been noted for its inaccuracy in predicting the dispersion minima \cite{Kreisel2009, Sonin2017}, it is deemed sufficient for the current analysis. The approximate dipole-exchange plane-wave dispersion is expressed through the relation:  
		\begin{align}
		\label{11}
		&\omega_n(k)^2 \approx (\omega_H +\omega_M \eta k_n^2)(\omega_H +\omega_M \eta k_n^2 + \omega_M \mathcal{K}),
		\end{align}
		where $\omega_H = \gamma_e B_z$ is the applied magnetic field strength, $k_n^2 = k^2 + k_{\perp}^2$ is the sum of the squared in-plane and perpendicular wavenumbers, and
		\small  
		\begin{align}
		\mathcal{K} &= P + \bigg(1-P+P\cos^2{\phi_B}+\frac{\omega_M(P-P^2)\sin^2{\phi}}{\omega_H +\omega_M \eta k_n^2}\bigg)\sin^2{\theta_B},\\ 
		P &= \frac{k^2}{k_n^{2}}-\frac{k^4}{k_n^{4}}\frac{2[1-(-1)^{n}e^{-kd}]}{kd(1 + \delta_n)},
		\end{align}
		\normalsize
		where $\theta_B$ is the angle between the magnetic field vector and the thin-film surface, and $\phi_B$ is the angle between the magnetic field vector and the wave propagation direction $k$. For thin films where the surface spins are described as being completely pinned or unpinned, and where the film thickness is in the order of the exchange field stiffness ($d\leq \sqrt{\eta}$), the perpendicular component of the wavenumber is quantized such that $k_\perp=\pi n/d$.     
		\begin{figure}
			\includegraphics[scale = 0.55]{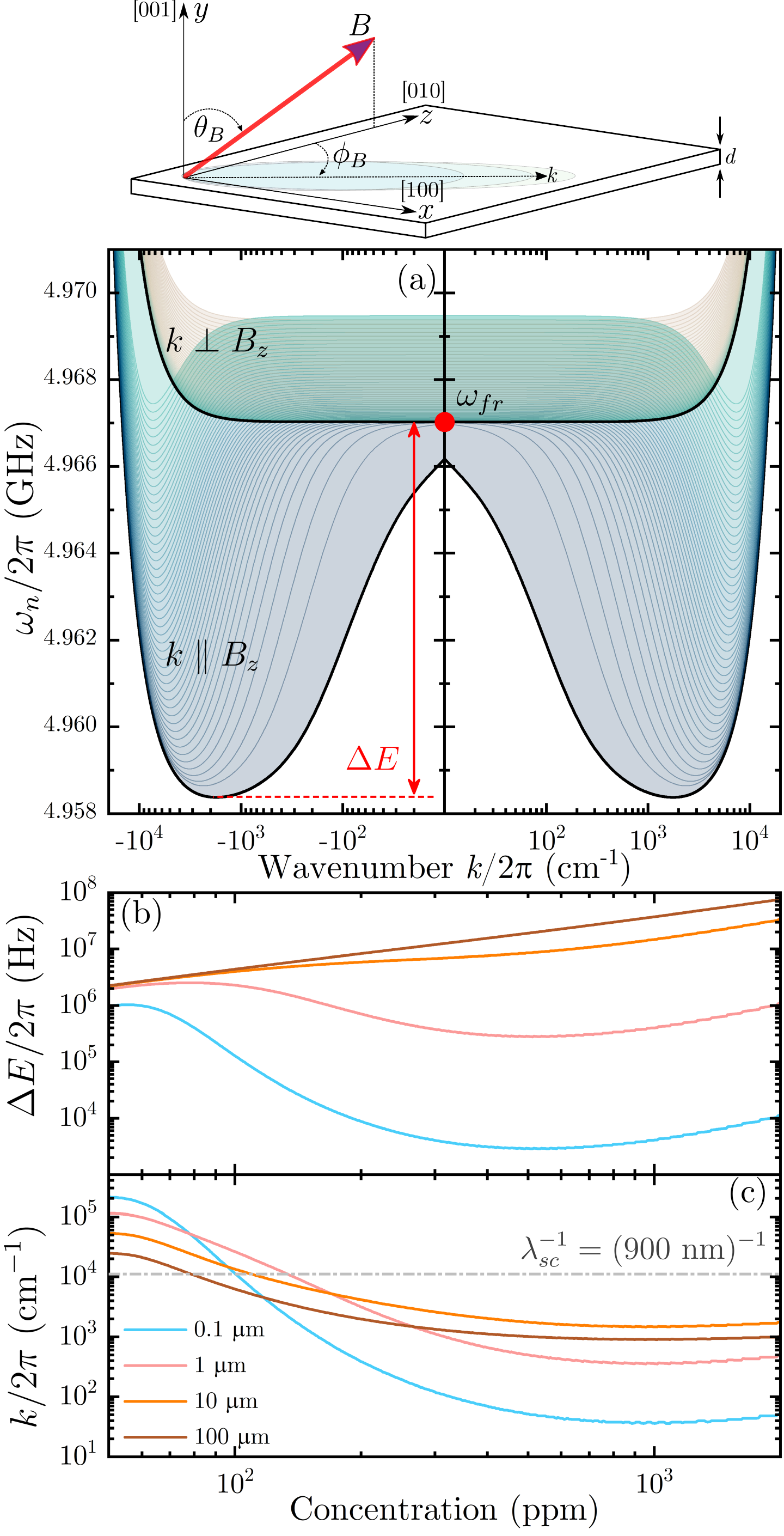}
			\centering
			\caption{(a) Spin-wave dispersion for the first 40 thickness modes in a 100-$\mu$m-thick diamond with a N-$V^-$ spin density of 200 ppm, highlighting the dispersion of plane spin-waves that propagate either parallel (longitudinal modes) or perpendicular (transverse modes) in relation to the externally applied magnetic field. (b) Dependence of the trap depth $\Delta E$ and (c) minimum $k$ values at the minimum energy, as a function of concentration and for varying thicknesses $d$.}
		\end{figure}
	
		The dipole-exchange spin-wave dispersion for a N-$V^-$ spin density of 200 ppm in a 100 $\mu$m thin diamond film is calculated using Eq.(\ref{11}) and plotted in Fig.5(a). It shows the dispersion of the first 40 longitudinal and transverse modes, for a magnetic field applied along the $[010]$ crystal axis. Both fundamental longitudinal and transverse branches (highlighted in black) converge to the ferromagnetic resonance energy $\omega_{fr} = \sqrt{\omega_H(\omega_H+\omega_M)}$. The distinctive minima of the dipole-exchange dispersion curves for longitudinal modes represents a clear condensation trap, for which its separation from $\omega_{fr}$ to the lowest dispersion energy by $\Delta E$ modifies the time and density needed for the formation of a condensate. The shape of the longitudinal and transverse branches highlight the two competing components in Eq.(\ref{11}), namely the exchange and dipole contributions, respectively, $\eta$ and $\omega_M$. For wavenumbers approaching the reciprocal lattice length, the dispersion increases quadratically as the interaction is dominated by the exchange energy. Similarly, for transverse mode spin-waves, the spin dipoles are aligned perpendicularly to the propagation direction, such that only the exchange interaction facilitates propagation, for which a quadratic $k_n$ relation dominates with a minimum at $k=0$. In the longitudinal branches, the dipole interaction dominates at wavenumbers smaller than approximately $\eta^{-1/2}$, resulting in a negative dispersion  which is subsequently exceeded by the exchange interaction when $k>\eta^{-1/2}$, such that at a finite $k$ a ``trap'' is formed. 
		
		Figures 5(b) and 5(c) highlight the trap depth $\Delta E$ and the wavenumber at the minimum dispersion energy, as a function of concentration. The trap depth increases as a function of concentration, but for thinner thicknesses there is a non-linear dependence which highlights the imbalance of the simultaneous increase of $\eta$ and $\omega_M$. For thinner layers, the rate of increase of $\omega_M$ is reduced and the longitudinal branch takes on a dominantly transverse shape, which becomes accentuated as the thickness decreases. In comparison, the YIG system displays a fixed wavenumber minima around $10^{-4}$ cm$^{-1}$, depending on the thin-film thickness \cite{Kalinikos1986a}, while the depth of the dispersion $\Delta E$ is in the order of gigahertz, due to the much higher saturation magnetisation. The change in the location of the minimum wavenumber as a function of concentration is also affected by the layer thickness, for the same reason, such that as the dispersion takes on a more transverse character, $k_{min}$ approaches zero. This value is crucial when considering experimental detection using angular-resolved Brillouin scattering, which for thin films is most conveniently measured using a back-scattering geometry. The detection window of such a configuration is limited by $k_{max} = 2(2\pi/\lambda_{sc})\sin{\theta}$, and the choice of wavelength is limited by the properties of the N-$V^-$ ensemble. In order to preserve the ground-state spin population and ensure that the charge state is minimally perturbed, $\lambda_{sc}$ needs to exceed the luminescence side-band and is therefore longer than approximately 850 nm \cite{Doherty2013}. The accessible wavenumber using a scattering laser wavelength of 900 nm is highlighted in Fig.5(c). Given this limitation, the minimum concentration needs to exceed 100 ppm for the detection of spin-waves in a back-scattering geometry to be feasible. Considering the likelihood of condensation, there is no minimum necessary concentration provided that the local temperature is below $T_c$; however, the longevity of condensation is strongly concentration and pump-dependent. 
		
	\subsection{Condensation Dynamics}
		
		The relaxation and condensation of magnons into their minimum energy state is facilitated by an inelastic scattering mechanism which conserves the number of magnons generated. Magnons will scatter with all crystal lattice irregularities, as well as with themselves in any number of combinations, so there are in principle many possible decay mechanisms. However, it has been argued \cite{Serga2014} that the most likely mechanism facilitating condensation is four-magnon scattering,	due to its total momentum and number conserving mechanism, and its greater occurrence likelihood. The main competing, non-number-conserving decay mechanism is spin-lattice relaxation ($\gamma$), and to a lesser degree, destructive scattering from lattice defects, although this is not considered here. The rudimentary analysis described here also does not consider the effects of dephasing, which is crucial when considering the paramagnetic/ferromagnetic resonance line-width and its spectral obfuscation of detecting condensation. It is, however, sufficient in providing an indication for the feasibility of condensation when considering the principle rates.     
		
		\begin{figure}
		\includegraphics[scale = 0.5]{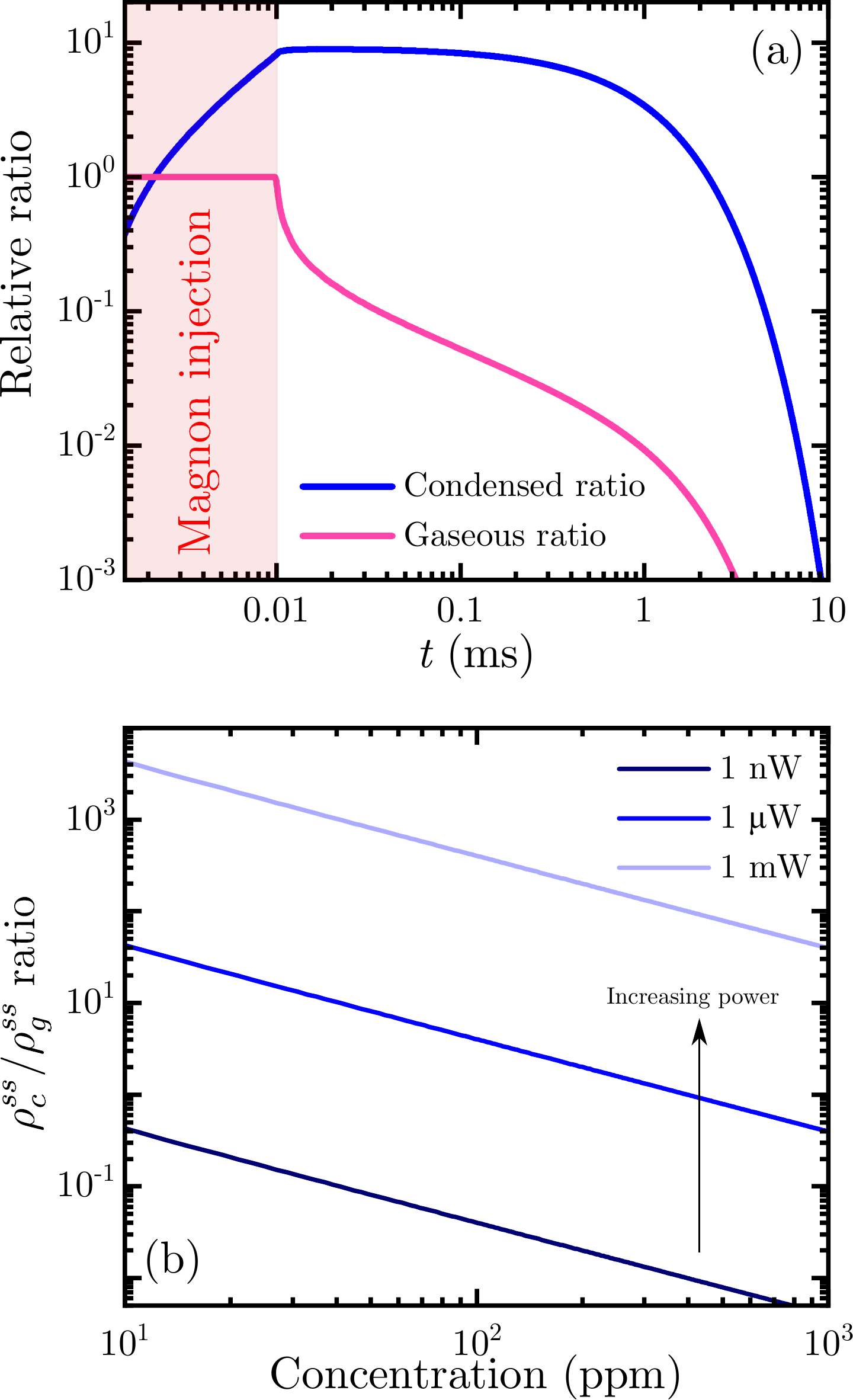}
		\centering
		\caption{(a) Decay of the total composition of magnons after 10 $\mu$s parametric pumping, for a concentration of 200 ppm, at $T=T_c \approx 5$ K and $\gamma/2\pi = (1~\text{ms})^{-1}$. (b) The steady-state ratio as a function of concentration and their equivalent $T_c$ for various pumping rates. As the concentration is increased, $T_c$ is increased, and the steady-state ratio is reduced. Conversely, as the pumping rate increases, the ratio is also increased due to the generation of more magnons, provided that $T<T_c$.}
		\end{figure}    
		
		Similar to the description in \cite{Serga2014}, a qualitative understanding of the magnon dynamics may be obtained using the following set of rate equations for an open two-level system:
		\begin{align}
		\label{14}
		\dot{\rho_c}(t) &= -\Gamma_p(t)-\gamma\rho_c + 2\bigg(\frac{\omega_{fr}}{\Delta E}\bigg)\xi \rho_g^3,\\
		\label{15}
		\dot{\rho_g}(t) &= 2\Gamma_p(t)-\gamma\rho_g - 2\bigg(\frac{\omega_{fr}}{\Delta E}-\frac{1}{2}\bigg)\xi \rho_g^3,
		\end{align}
		where $\rho_{c/g}$ is the population of the condensed or gaseous portion of the parametrically injected magnons, $\Gamma_p(t)$ is the time-dependent parallel parametric pumping rate which is assumed to be perfectly efficient, and $\xi$ is the four magnon scattering rate. The parallel parametric pumping mechanisms of spin-waves is an inherent phenomenon \cite{Schlomann1960} by which a spin-wave mode is excited by a frequency-doubled microwave that is applied along the magnetisation direction. Due to the momentum conservation law, this creates two spin-waves with equal but opposite wavenumber, at half the energies of the applied field.       
		
		The general steps toward condensation considered here are instigated by the injection of magnons at a frequency of $2\omega_{fr}$, which decay towards $\omega_{fr}$ via the previously described parametric instability. These magnons form a gas with energies that span from the lowest dispersion energy to a maximum which exceeds $\omega_{fr}$ but is much lower than the continuum of magnon energies that are equilibrated with the surrounding temperature. The nonlinear four-magnon scattering involves two magnons exchanging their energy while conserving their total momentum, and results in one magnon decaying into a lower energy state and another into a higher state. Here the likelihood of one magnon decaying into the condensate is regulated by the $\omega_{fr}/\Delta E$ ratio, while the second magnon always escapes the gaseous fraction into the room-temperature continuum. With a deeper trap, and a smaller $\omega_{fr}/\Delta E$ ratio, the magnon gas occupies a larger span of energies, thereby reducing the likelihood for scattering to occur directly into the condensate energy state for a given magnon injection rate. 
		
		These rate equations are set with respect to the parameters extracted from the calculated dispersion curves in the previous section, where the first terms represent the injection/removal of gaseous/condensed magnons when the pump is applied, the second term represents the destruction of magnons through the spin-lattice decay rate, and the third and fourth terms represent the nonlinear scattering of the magnons into either the condensed fraction or out of the system. The spin-lattice relaxation rate for a N-$V^-$ ensemble is reported to be in the order of kilohertz (1-10 ms \cite{Jarmola2012}), and is set here to be $\gamma/2\pi = (1~\text{ms})^{-1}$, while the four-magnon scattering rate may be approximated as a function of concentration-based $T_c$ and the local temperature $T$ \cite{Cornelissen2016}:
		\begin{align}
		\xi \approx \bigg(\frac{k_B T}{h}\bigg)\bigg(\frac{T}{T_c}\bigg)^3,
		\end{align}
		which spans from the hundreds of terahertz to the tens of gigahertz for concentrations between 100 and 1000 ppm. Comparatively, YIG samples have been reported to posseses four-magnon scattering rates in the order of 10 to 100 GHz, relative to its spin-lattice relaxation rate in the order of 2 MHz.              
		
		An example of the temporal dynamics is plotted in Fig.6(a) in terms of the relative population after a 10 $\mu$s window of sustained pumping at a pump power and frequency of 1 $\mu$W and 10 GHz, respectively. A consequence of the large $\xi/\gamma$ ratio is the relative magnitude of the magnon condensate. When further examining the relative concentration in the steady-state (see Appendix \ref{C}) in Fig.6(b), there is an observed decrease in the relative condensate fraction as the concentration of spins is increased. As the average exchange energy increases, $\omega_{fr}$, and therefore the trap depth, increases along with the distribution of magnon energies. For a given injection rate, a smaller fraction of the magnon gas is likely to condense compared to those that escape. This is counteracted by increasing the magnon-injection rate, in order to ensure that the pump generates more magnons, and a comparatively larger portion of the gas remains condensed, provided that $T<T_c$..
		
		\subsection{Detection of a N-$V^-$ Magnon Condensate} 
		
		Given the predictions of the analysis presented so far, considering a conceptual experiment to detect magnon condensation in an N-$V^-$ ensemble is warranted. An obvious detection scheme, identical to that used for YIG systems, is Brillouin scattering. As a dense N-$V^-$ ensemble is expected to render the diamond crystal opaque, reflection-based scattering is considered. A diamond crystal which has been thinned down to a thickness below 100 $\mu$m, and contains a dense N-$V^-$ ensemble is first subjected to continuous excitation at around 532 nm for a given duration. The duration of this pulse depends on the spin density and applied power, and ensures that the spins are initialised into their $\vert0\rangle$ ground-state. This is normally an up-to 80 \% efficient procedure for single spins at room temperature \cite{Robledo2011}, but will be less efficient when applying an off-axial magnetic field due to the spin-mixing. This is followed by applying a microwave field using a e.g. co-planar wave-guide, aligned in parallel with the external magnetic field which is aligned along the $[010]$ crystallographic axis, in order to generate plane spin waves that propagate in parallel with the magnetic field. A probe beam is then shone onto the diamond either while the microwave is continuously generated to observe the steady-state spin-wave density, or immediately after the microwave is stopped to observe the temporal dynamics. As mentioned in Sec. III.A, the wavelength of the probe beam needs to exceed the luminescence side-band of the N-$V^-$ centre in order to ensure that ground-state spin population is minimally perturbed. This necessitates wavelengths beyond 850 nm, and possibly even exceeding 1050 nm in order to avoid any finite absorption via the singlet-state transition of the N-$V^-$ system. 
		
		If propagating spin waves are generated, a portion of the back-scattered  light will be Stokes-shifted with the energy of the spin wave, and detectable through a high-resolution spectrometer such as a variable-length Fabry-Perot cavity. The amplitude of the Stokes-shifted signal should be proportional both to the applied probe-field power, and to the density of spin-waves present. The amplitude of the Stokes-shifted peak will therefore highlight the minima of the dispersion when measured as a function of the applied microwave frequency, as the magnons are expected to scatter down toward their lowest energy, irrespective of whether they form a condensate or not. The determination of the formation of a condensate will be witnessed through the time-dependent amplitude of the Stokes-shifted peak, whose variation will highlight the onset and decay of condensation. 
		
		With the calculated $T_c$ values in mind, and the limitation of the minimum wavenumber highlighted in Fig5(c), a minimum concentration of approximately 100 ppm is necessary. Given the technical challenges of obtaining concentrations that exceed even just 10 ppm, preliminary experiments are expected to be carried out with concentrations approaching 100 ppm, which necessitates the use of an optically accessible dilution fridge to reach base temperatures below 1 K. This places limitations on the injected optical and microwave power in order to maintain a stable base temperature. For concentrations exceeding 200 ppm, simpler cryostats with base temperature exceeding 2 K may be used with less stringent limitations on the injected probe light and microwave field.	 		
		
\section{Conclusion}

		The formation of a magnon condensate in a N-$V^-$ ensemble is explored here by considering the magnetic order of a randomly distributed spin ensemble. This was carried out by estimating the density-dependent dipole and exchange energy using a rudimentary $sp^3$ hybrid description of the N-$V^-$ spin orbitals. The calculated spread of energies was parameterised as a function of concentration through assessing the corresponding average nearest-neighbour distance, which was then used in the numerical calculation of the magnetic phase transition temperature $T_c$. Below $T_c$, spatial randomness was assumed to be a negligible perturbation, and employing the standard description for dipole-exchange spin-wave dispersion was justified. The spin-wave dispersion was calculated using the estimated exchange energies, in order to extract concentration dependent dispersion parameters, in particular the depth of the dispersion trap, and the corresponding wavenumber at the lowest dispersion energy. This was followed by using the extracted parameters in combination with $T_c$ and a measured spin-lattice relaxation rate to assess the relative size and longevity of a magnon condensate compared to its gaseous state. Ultimately, it is postulated that a magnon condensate may be observable for concentrations exceeding 90 ppm at low temperatures, and approximately 450 pm at room temperature through Brillouin back-scattering.       
		
		The analysis presented here is by no means thorough; further refinement is necessary in the choice of wavefunction and the Hamiltonian used to predict the onset of magnetic order, while the crucial role of spin dephasing needs to be accounted for. However, in pursuit of novel systems for the study of macroscopic quantum phenomenon, this preliminary study highlights an intriguing possibility in using the N-$V^-$ system, which has already proven to be of such convenient and practical importance in the study of quantum information and metrology. Although the current physical limitations on creating dense spin ensembles in diamond are considerable, they are expected to be overcome as the understanding of material growth is further advanced. In light of this, the generation of concentrations approaching 100 ppm is conceivably within the grasp of current irradiation technology and epitaxial growth, and an abundance of unique physics is anticipated when concentrations approach values large enough to incur magnetic self-ordering.   
		
\begin{acknowledgments}
		This work is lovingly dedicated to the memory of Malika Ben Moussa. Funding from the Villum Foundation (grant No.17524) is gratefully acknowledged. 
 
\end{acknowledgments}

\appendix
		
		\section{Spin Wavefunctions} \label{A}
		The orbitals of the N-$V^-$ centre are described using $sp^3$ hybridised orbitals defined for the nitrogen defect ($n_1$) and the three carbon atoms surrounding the vacancy defect ($c_1,c_2,c_3$):
		\begin{align}
		n_1 &= \frac{1}{2}\psi_{\text{N},2s} - \frac{\sqrt{3}}{2}\psi_{\text{N},2p_z},
		\\
		c_1 &= \frac{1}{2}\psi_{\text{C},2s}+\frac{1}{2\sqrt{3}}\psi_{\text{C},2p_z} + \sqrt{\frac{2}{3}}\psi_{\text{C},2p_y},	
		\\
		c_2 &= \frac{1}{2}\psi_{\text{C},2s}+\frac{1}{2\sqrt{3}}\psi_{\text{C},2p_z}-\frac{1}{\sqrt{6}}\psi_{\text{C},2p_y} - \frac{1}{\sqrt{2}}\psi_{\text{C},2p_x},
		\\
		c_3 &= \frac{1}{2}\psi_{\text{C},2s}+\frac{1}{2\sqrt{3}}\psi_{\text{C},2p_z}-\frac{1}{\sqrt{6}}\psi_{\text{C},2p_y} + \frac{1}{\sqrt{2}}\psi_{\text{C},2p_x},
		\end{align}  
		where the single orbital wave functions are defined using single-electron hydrogen wave functions in the form $\psi_{Z,n,l,m}(r,\theta,\phi) = \text{R}_{n,l}(r)\text{Y}_{l,m}(\theta,\phi)$, where $\text{R}_{n,l}(r)$ is the normalised Laguerre polynomial and $\text{Y}_{l,m}(\theta,\phi)$ is the spherical harmonic function. The normalised N-$V^-$ orbital wave functions are then defined through \cite{Lenef1996, Hossain2008}:
		\begin{align}
		u &= n_1,
		\\
		v &= \frac{c_1+c_2+c_3-3\langle c_1 \vert u \rangle u}{\sqrt{3(1+2\langle c_3 \vert c_2 \rangle-3\langle c_1 \vert u \rangle^2)}},
		\\
		e_x &= \frac{2 c_3-c_2-c_1}{\sqrt{3(2-2\langle c_3 \vert c_2 \rangle)}}, 
		\\
		e_y &= \frac{c_2-c_3}{\sqrt{2-2\langle c_3 \vert c_2\rangle}}.
		\end{align}  
		Using these expressions, the integrals that define the exchange constant $J_{ex}$ [Eq.(\ref{5})-(\ref{7})] are numerically solved using Monte Carlo integration, where the error bound is inversely proportional to the square root of the number of trials. 
		
		\section{Sherrington-Kirkpatrick Model} \label{B}
		Of the various iterations of the spin Ising model which account for randomness, the Sherrington-Kirkpatrick model is one of the most studied so far. The typical Hamiltonian [Eq.(\ref{8})] describes the competing strengths between long-range spin order and an extrinsic magnetic field, which may both be described in terms of a random field probability distribution function. In the Gaussian case:
		\begin{align}
		\label{B1}
		P(W_{ij}) &= \frac{e^{-(W_{ij}-W_0)^2 /2 W^2 }}{\sqrt{2\pi} W },\\
		\label{B2}
		P(h_i)    &= \frac{e^{-(h_{i}-h_0)^2/2\Delta^2 }}{\sqrt{2\pi}\Delta}.
		\end{align}	
		The main objective is in extracting an expression for the free energy of the system and the variously attributed magnetic order parameters in the thermodynamic limit such that $N\rightarrow\infty$. There are no known analytical solutions in finite dimensions, and numerical solutions are intractable, as the computational duration is impractically long for the averaging of all possible arrangements of a randomly distributed \textit{N}-sized system. Heuristic methods have thus been developed, such as the `replica-trick', which uses the logarithmic limit:
		\begin{align}
		\label{B3}
		\ln x = \lim_{n\rightarrow 0} \frac{x^n -1}{n}. 
		\end{align}     
		The free energy $\mathcal{F}$ in the thermodynamic limit is expressed as:
		\begin{align}
		\label{B4}
		\mathcal{F}(\beta) = \lim_{N\rightarrow \infty} \frac{1}{N\beta}\ln Z(\beta,N), 
		\end{align}     
		where $Z(\beta,N)$ is the canonical partition function. Randomness is accounted for by taking the thermal average over the random disorder such that:
		\begin{align}
		\label{B5}
		\mathcal{F}(\beta) = \lim_{N\rightarrow \infty} \frac{1}{N\beta}\big\langle\ln Z(\beta,N)\big\rangle. 
		\end{align}
		The direct calculation of the partition function in this case is intractable, and the identity Eq.(\ref{B3}) is used which is reinterpreted to imply that there are $n$-configurations or `replicas' of the system such that $Z^n = \prod^n_{\alpha=1}Z_\alpha$, thereby reformulating Eq.(\ref{B5}) into the solvable expression:
		\begin{align}
		\label{B6}
		\mathcal{F}(\beta) = \lim_{n\rightarrow 0}\lim_{N\rightarrow \infty} \frac{1}{nN\beta}\big(\langle Z^n(\beta,N)\rangle-1\big). 
		\end{align}
		Using the Hubbard-Stratanovich transformation, an expression for the free energy per particle is obtained in the set limits with respect to the magnetic order parameter $m$ and the Edward-Anderson order parameter $q$:
		\begin{align}
		q &= \int^{\infty}_{-\infty}\frac{e^{-z^2/2}dz}{\sqrt{2\pi}}\text{tanh}^2(\zeta),\\
		m &= \int^{\infty}_{-\infty}\frac{e^{-z^2/2}dz}{\sqrt{2\pi}}\text{tanh}(\zeta),\\\nonumber
		\zeta &= \beta \big[h_0 + W_0 m + z W \sqrt{q + (\Delta/W)^2}\big],\nonumber 
		\end{align}			  		
		from which the critical transition temperature $T_c$ of Eq.(\ref{9}) can be derived for the condition $m \neq 0 $. A thorough exposition can be read in \cite{Binder1986,Hadjiagapiou2018}.
		
		\section{Coupled Rate Equations and Steady-State Solution} \label{C}
		The derivation of the rate equations Eq.(\ref{14}) and (\ref{15}) is based on the steps and assumptions presented in \cite{Serga2014}. The initial assumption is that the gaseous magnons are in thermodynamic equilibrium with the energy of the injected magnons $E_{in}$, and that they span an energy range which is much lower than the energies of room-temperature magnons, such that $k_BT\gg E_{max}>E>E_{min}$, where $E_{min} = \hbar\omega_{fr} - \Delta E$, and $E_{max}$ is an upper cut-off energy which is temperature independent. Assuming that a condensate has already formed, the total number of magnons present in the system is the sum of the number of gaseous and condensed magnons, $\rho_g$ and $\rho_c$, respectively, and is related to the total energy such that:
		\begin{align}
		\label{C1}
		E_{tot} = E_{gas} + E_{cond} = E_g\rho_g + E_{min}\rho_c,   
		\end{align}
		where $E_g$ is the average energy of the gaseous fraction, defined in terms of the magnon integrated density of states in the low-energy regime (Rayleigh-Jeans condition):
		\begin{align}
		\label{C2}
		E_g &= \frac{\int^{E_{max}}_{E_{min}}EdN_g}{\int^{E_{max}}_{E_{min}}dN_g}.\\
		\label{C3}
		dN_g &= \frac{\sqrt{2}Vm_e^{3/2}}{\hbar^2\pi^2}\frac{k_BT}{\sqrt{E-E_{min}}} dE,
		\end{align}
		where $V$ is the volume, and $m_e$ is the effective mass of the magnon. Equation (\ref{C3}) is derived with the assumption that a condensate already exists and the chemical potential $\mu$ is equal to the condensates energy which is the minimum energy of the dispersion, $\mu = E_{min}$. To simplify the rate equations, the expression in Eq.(\ref{C2}) can be approximated to $E_g \approx\hbar\omega_{fr}-\tfrac{1}{2}\Delta E$ with an appropriate choice of $E_{max}$, which can be arbitrarily chosen as long as $E_{in}<E_{max}\ll k_BT$. Here it is set to $E_{max}\approx\hbar\omega_{fr}+E_g$. Using these relations, the rate equations (\ref{14}) and (\ref{15}) are derived by initially defining them in terms of the total energy and number of gaseous magnons, and then rearranging to obtain the final expressions, as detailed in \cite{Serga2014}. Their steady-state solutions are then obtained through rearrangement to obtain a cubic expression in terms of the gaseous fraction $\rho_g^{ss}$:
		\begin{align}
		\label{C4}
		\rho_c^{ss} &= \bigg[-\Gamma_p + 2\bigg(\frac{\omega_{fr}}{\Delta E}\bigg)\xi\rho_g^{_{ss}3}\bigg]\gamma^{-1},\\
		\label{C5} 
		\rho_g^{ss} &= \bigg[2\Gamma_p -2\bigg(\frac{\omega_{fr}}{\Delta E}-\frac{1}{2}\bigg) \xi\rho_g^{_{ss}3}\bigg]\gamma^{-1},
		\end{align}
		for which the steady-state gaseous fraction is obtained through the single real solution of the cubic equation (\ref{C5}).
\bibliography{Biblio}
\end{document}